\documentclass[superscriptaddress,floatfix,reprint,amsmath,amssymb,aps,prl]{revtex4-2}

\usepackage[hidelinks]{hyperref}
\usepackage[capitalize]{cleveref}
\usepackage{physics}
\usepackage{xcolor}
\usepackage{xspace}
\usepackage{float}
\usepackage{mathtools}
\usepackage{titlesec}

\newcommand{\mysection}[1]{\textit{#1.---}}

\renewcommand{\d}{\mathrm{d}}
\newcommand{\ud}{\mathrm{ud}}
\newcommand{\phasediff}{(\phi_\d - \phi_\ud)}
\newcommand{\averagephase}{\bar \phi}

\begin{document}
\date{May 7, 2025}

\newcommand{\mytitle}{Nonreciprocal interactions induce frequency shifts in superradiant lasers}

\title{\mytitle}

\author{Tobias Nadolny}
\affiliation{Department of Physics, University of Basel, Klingelbergstrasse 82, 4056 Basel, Switzerland}
\author{Matteo Brunelli}
\affiliation{JEIP, UAR 3573 CNRS, Coll\`ege de France, PSL Research University, 11 Place
Marcelin Berthelot, 75321 Paris Cedex 05, France}
\author{Christoph Bruder}
\affiliation{Department of Physics, University of Basel, Klingelbergstrasse 82, 4056 Basel, Switzerland}

\begin{abstract}
\normalsize
Superradiant lasers,
which consist of incoherently driven atoms coupled to a lossy cavity,
are a promising source of coherent light due to their stable frequency and superior narrow linewidth.
We show that when a fraction of the atoms is not driven, a shift in the lasing frequency and a broadening of the linewidth occur, limiting the performance of a superradiant laser.
We explain this behavior by identifying nonreciprocal interactions between driven and undriven atoms, i.e., competing alignment and antialignment of their dipoles.
Our results have implications for the realization of superradiant lasers, establishing the relevance of nonreciprocal phenomena for quantum technologies.
\end{abstract}

\maketitle

Superradiant lasers harness the coherence of a large ensemble of atoms that are collectively coupled to a cavity mode to provide light with narrow linewidth~\cite{Haake_1993,Chen_2009,Meiser_2009}.
In contrast to a standard laser, the cavity decays rapidly ensuring a small number of excitations in the cavity mode, which renders the laser robust against cavity fluctuations~\cite{Bohnet_2012,Norcia_2016}.
The laser light exhibits a stable frequency set by the atomic transition frequency.
Continuous lasing can be achieved by incoherently driving each atom to provide it with energy~\cite{Norcia_2016,Laske_2019,Kristensen_2023}.
Superradiant lasers offer great technological promise as their exceptionally narrow linewidth is expected to significantly improve the precision of optical atomic clocks~\cite{Ludlow_2015}.
\begin{figure}[t]
    \centering
    \includegraphics{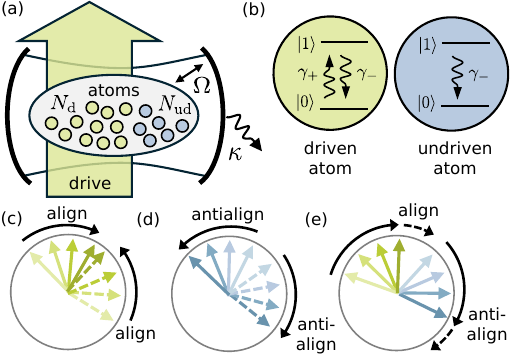}
    \caption{
    (a) Superradiant laser, where $N_\d$ atoms are driven and $N_\ud = N-N_\d$ are undriven.
    All atoms are coupled at rate $\Omega$ to the cavity which decays at rate $\kappa$.
    (b) Incoherent gain and loss processes within each atomic spin.
    (c) Alignment between two driven atomic dipoles (solid and dashed straight arrows). Light to darker color indicates increasing time. 
    (d)~Antialignment between two undriven dipoles.
    (e)~Nonreciprocal interactions between driven and undriven dipoles.
    The dashed circular arrows indicate the continuous motion.
    }
    \label{fig:1}
\end{figure}

In this letter, we study a superradiant laser where a fraction of the atoms is \textit{not} driven; see Figs.~\ref{fig:1}(a,b).
Surprisingly, this modification results in a shift of the lasing frequency and spectral broadening, which may be detrimental to the
use of a superradiant laser as a stable frequency reference with narrow linewidth.
This contrasts with the expectation that the undriven atoms behave as passive spectators, causing only a reduction of the laser power.
To explain this result, we show that driven and undriven atoms interact in a nonreciprocal way.

Nonreciprocal interactions mean that entities influence each other in asymmetric ways.
They are an exciting feature of active matter~\cite{Bowick_2022,Shankar_2022}, i.e., an ensemble of interacting active agents, resulting in nonreciprocal phase transitions~\cite{You_2020,Fruchart_2021,Brauns_2024} and emergent time-crystalline order~\cite{Hanai_2024}.
Here, we consider the superradiant laser as quantum active matter and the atomic dipoles as quantum active agents that can interact in a nonreciprocal way.
We find that the driven atoms tend to align their dipole with the dipoles of both driven and undriven atoms;
the undriven atoms, however, have the opposite inclination: they tend to align their dipoles opposite to those of all other atoms; see Figs.~\ref{fig:1}(c-e).
The competition of alignment and antialignment results in chase-and-run-away dynamics of the atomic dipoles, which translates to a frequency shift of the laser. 
The antialignment itself decreases the coherence among the atoms, causing spectral broadening and loss of power.

Our work connects fundamental concepts of active matter with quantum technologies by showing that nonreciprocal interactions have practical implications for the development of superradiant lasers.
Furthermore, our model offers a physical realization of conformist-contrarian dynamics,
introduced in Refs.~\cite{Hong_2011_PRL, Hong_2011_PRE} inspired by social behavior, whose implementation remained unclear. 
More broadly, our findings contribute to the understanding of nonreciprocal phenomena in driven-dissipative quantum systems~\cite{Hanai_2019,Hanai_2020,Sieberer_2023,Zelle_2024} beyond unidirectional interactions~\cite{Gardiner_1993,Carmichael_1993,Metelmann_2015}.
While recent studies have shown that nonreciprocal interactions can be engineered by carefully designed light-matter couplings~\cite{Chiacchio_2023,Reisenbauer_2023,Rudolph_2024,Hanai_2024_arxiv,Nadolny_2024},
the present work demonstrates that quantum nonreciprocal interactions can naturally occur between driven and undriven atoms in a superradiant laser.

\mysection{Model}%
We describe a superradiant laser, which comprises $N$ atoms each modeled as a quantum spin-1/2 with states $\ket{0}$ and $\ket{1}$; see \cref{fig:1}.
While all spins are coupled coherently to a lossy cavity mode, only $N_\d \leq N$ of the spins are incoherently driven; the other $N_\ud =  N - N_\d$ spins are not driven.
The fraction of driven (undriven) spins is $p_\mathrm{d} = N_\d/N$ ($p_\ud = 1-p_\d$).
The spins are described by Pauli matrices $\sigma^z_{\mu,i} = \dyad{1}_{\mu,i} - \dyad{0}_{\mu,i}$ and ladder operators $\sigma^+_{\mu,i} = \dyad{1}{0}_{\mu,i}$, $\sigma^-_{\mu,i} = \dyad{0}{1}_{\mu,i}$.
The index $\mu \in \{\d,\ud\}$ distinguishes driven and undriven spins, and $i$ ranges from $1$ to $N_\mu$ .
The collective spin operators are $S^\pm = \sum_{i=1}^{N_{\d}} \sigma_{\d,i}^\pm + \sum_{i=1}^{N_{\ud}} \sigma_{\ud,i}^\pm$.

The quantum Lindblad master equation for the density operator $\rho$ is
\begin{equation}
    \dot \rho
    =
    -i\Omega[a^\dag S^- + a S^+,\rho] + \kappa \mathcal{D}[a]\rho
    +\gamma_+ \sum_{i=1}^{N_\d} \mathcal{D}[\sigma^+_{\d,i}]\rho 
    \, ,
	\label{eq:master}
\end{equation}
where $\mathcal{D}[o]\rho = o\rho o^\dag - (o^\dag o \rho + \rho o^\dag o )/2$ is the Lindblad dissipator.
Both the atomic spins and the cavity are described in the frame rotating with their bare frequencies, which are set equal for simplicity.
A nonzero spin-cavity detuning does not influence our results, see Supplemental Material~\cite{supp}.
All spins couple equally with strength $\Omega$ to the cavity mode $a$, which decays at a rate $\kappa$.
The last term in \cref{eq:master} describes an incoherent drive at rate $\gamma_+$, which can be engineered by pumping the state $\ket{0}$ to a third state which rapidly decays to state $\ket{1}$.
For the case of all atoms being driven, $N_\d = N$, the master equation reduces to the model studied in Ref.~\cite{Meiser_2009}.

\mysection{Emission spectrum}%
A key quantity to characterize a laser is the spectral density $S(t,\omega)$ of light emitted by the cavity at time $t$ and per frequency $\omega$,
\begin{equation}
    S(t,\omega) = 
    \int_{-\infty}^\infty \mathrm{d} \tau \expval{a^\dag(t+\tau)a(t)} e^{i\omega\tau}
    \, .
\end{equation}
The steady-state spectrum is defined as $S(\omega) = \lim_{t\rightarrow \infty} S(t,\omega)$.
The spectrum of a high-quality laser comprises a large and narrow peak, implying high output power and small linewidth.
The spectrum can be calculated by employing a cumulant expansion approximation and the quantum regression theorem~\cite{cumulants_kubo,Breuer,Meiser_2009,Xu_2013,Plankensteiner2022quantumcumulantsjl,supp}.
The approximation amounts to neglecting third and higher-order correlations.

\begin{figure}
    \centering
    \includegraphics[width =3.4in]{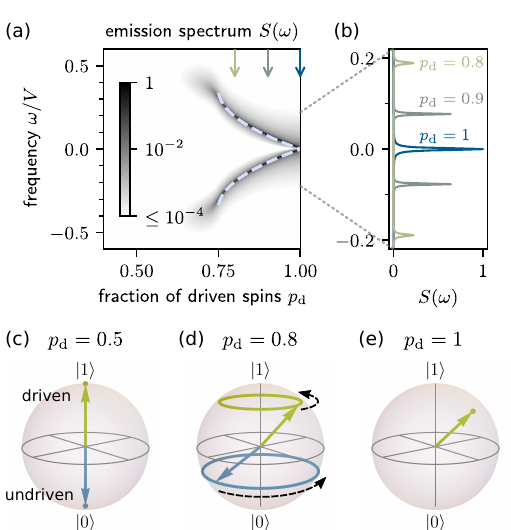}
    \caption{
    (a) Cavity emission spectrum $S(\omega)$ (arb.\ units) as a function of fraction $p_\d = N_\d/N$ of driven spins for $N=10^3$, $\kappa = 10 \sqrt{N} \Omega$, and $V=\gamma_+$.
    The frequency $\omega$ is measured relative to the bare atomic frequency.
    The dashed line indicates the mean-field frequency.
    Below the lasing transition $p_\d < 3/4$ [see \cref{eq:Vc_simple}], the emission is close to zero.
    (b) Line cuts through (a) as marked by the three arrows in (a) on a linear scale.
	(c-e)
	Solutions of the mean-field equations~\eqref{eq:MF_simple} in the long-time limit displayed on the Bloch sphere for
	(c) $p_\d = 0.5$: no lasing,
	(d) $p_\d = 0.8$: lasing with frequency shift, and
	(e) $p_\d = 1$: standard superradiant laser. 
    In (d), the dashed curved arrows indicate the continuous oscillations.
    }
    \label{fig:spectrum_pA}
\end{figure}

Figures~\ref{fig:spectrum_pA}(a,b) show the steady-state spectrum for the model defined in Eq.~\eqref{eq:master} as a function of $p_\d$.
If all spins are driven, $p_\d=1$, they collectively emit photons in a superradiant way into the cavity mode, which outputs highly coherent light as indicated by the narrow peak in the spectrum at zero frequency (relative to the bare atomic frequency).
When decreasing $p_\d$, there are two peaks in the spectrum at nonzero frequencies indicating a positive or negative shift in the lasing frequency.
Furthermore, for $p_\d$ smaller than a critical value, the peaks vanish indicating the absence of lasing.

\mysection{Nonreciprocal interactions}%
We now explain the origin of the frequency shift displayed in Figs.~\ref{fig:spectrum_pA}(a,b).
First, we derive an effective spin-only description.
In the bad-cavity limit where $\kappa/(\sqrt{N} \Omega)$ is large, the cavity field instantaneously follows the spins, $a = -i(2\Omega/\kappa)S^-$, and can be adiabatically eliminated.
Next, we exploit the permutational invariance: All driven spins are identical to one another and all undriven spins are identical to one another.
In the thermodynamic limit, $N \rightarrow \infty$, \cref{eq:master} can thus be solved exactly using a mean-field ansatz~\cite{Spohn_1980,Note1}.%
\footnotetext[1]{
The thermodynamic limit $N \rightarrow \infty$ is well defined, when simultaneously decreasing $\Omega \rightarrow 0$, such that $V \propto N\Omega^2 = \mathrm{constant}$
}
Introducing the coherences $s^+_{\mu} = \expval{\sigma^+_{\mu,i}}$, populations $s^z_{\mu} = \expval{\sigma^z_{\mu,i}}$, and the
average coherence $s^+ = p_\d {s}_{\d}^+ +p_\ud s_\ud^+$,
the mean-field equations are
\begin{equation}
    \begin{split}
  &\dot {s}_{\d}^+
  =
  V  s^+ s^z_{\d} 
  -\gamma_+ s^+_{\d}/2
\, ,
\\
  &\dot {s}_{\d}^z
  =
  -4V \Re[{s}_{\d}^- s^+]
  +
  \gamma_+\left( 1- {s}_{\d}^z \right)
  \, ,
  \\
  &\dot {s}_{\ud}^+
  =
  V  s^+ s^z_{\ud} 
\, ,
 \quad
  \dot {s}_{\ud}^z 
  =
  -4V \Re[{s}_{\ud}^- s^+]
  \, .
    \end{split}
    \label{eq:MF_simple}
\end{equation}
Here, $V = 2N\Omega^2/\kappa$ is the effective dissipative coupling strength among all spins.
Since we have not yet included any decay or dephasing of the spins, the purity of the undriven spins is conserved, $(s^z_\ud)^2 + 4s^+_\ud s^-_\ud = \mathrm{const} \leq 1$.

Equations~\eqref{eq:MF_simple} imply that the coherence of driven (undriven) spins tends to align in (out of) phase with respect to the average coherence.
This becomes explicit in the dynamical equations for the phases $\phi_{\mu} = \arg(s^+_{\mu})$ and the average phase $\averagephase = \arg(s^+)$ derived from \cref{eq:MF_simple}
\vspace{-0.23em} % this makes it fit on 4 pages :)
\begin{align}
    \dot \phi_{\mu} =
    s^z_{\mu} \frac{V \abs{s^+}}{\abs*{s^+_{\mu}}} \sin(\averagephase - \phi_{\mu}) \, .
    \label{eq:mf_sync}
\end{align}
The sign of the term multiplying the sine determines whether $\phi_{\mu}$ aligns or antialigns with $\averagephase$.
In a standard superradiant laser, all spins are driven, and their population is inverted $s^z_\d > 0$.
Consequently, the interactions are reciprocal: all spins tend to align their phases, resulting in a synchronized state (see \cref{fig:spectrum_pA}(e)) and the narrow linewidth of the emitted light~\cite{Meiser_2009,Zhu_2015}.
The population of the undriven spins, however, is not inverted and $s^z_\ud < 0$ is obtained in the long-time limit.
Therefore, the phase $\phi_{\ud}$ of the undriven spins is repelled from the average phase $\averagephase$.
The phase $\phi_{\d}$ of the driven spins remains to be attracted by the average phase $\averagephase$.
The competing attraction and repulsion of the phases constitute the effective nonreciprocal interactions among the atomic dipoles, summarized in \cref{fig:1}(b).

Dynamics of the form of \cref{eq:mf_sync} have been previously described for a network of classical phase oscillators~\cite{Hong_2011_PRL,Hong_2011_PRE}.
The oscillators that tend to align with the mean field have been termed conformists.
The oscillators that tend to antialign with the mean field oppose the average coherence; therefore, they have been termed contrarians.
The conformist-contrarian dynamics have been connected to opinion forming, but no physical model to realize them is proposed.
We have shown that the same phase interactions emerge in a superradiant laser with a fraction of undriven spins, offering a physical realization of these dynamics.
In Refs.~\cite{Hong_2011_PRL,Hong_2011_PRE}, the nonreciprocity results in chase-and-run-away dynamics of the phases, named traveling-wave states.
We now show that Eqs.~\eqref{eq:MF_simple} also result in traveling-wave states for $p_\d<1$, which implies emergent oscillations explaining the shift in the lasing frequency displayed in Figs.~\ref{fig:1}(a,b).

\mysection{Traveling-wave states}%
The mean-field equations Eqs.~\eqref{eq:MF_simple} can be solved exactly by an ansatz with constant populations $s^z_{\d,\ud}$, and oscillating
$s^+_\d = \abs{s^+_\d} e^{i\omega t + i \phasediff}$ and
$s^+_\ud = \abs{s^+_\ud} e^{i\omega t}$.
We introduced the shared oscillation frequency $\omega$, and the constant phase difference $\phasediff$ between driven and undriven spins.
In the Bloch-sphere picture, the ansatz describes oscillations on circles with radii $\abs{s^+_\mu}$ at constant $s^z_\mu$; see \cref{fig:spectrum_pA}(d).
Inserting this ansatz in Eqs.~\eqref{eq:MF_simple},
the frequency of the traveling-wave states is obtained as
\begin{equation}
        \omega = \pm \sqrt{\frac{\gamma_+}{4} \left(
         v - 2V p_\mathrm{ud}
        - \sqrt{v(v-4V p_\mathrm{ud})} \right)}
        \, ,
        \label{eq:omega}
\end{equation}
where $v = 2V-\gamma_+$.
The frequency is shown by the dashed line in \cref{fig:spectrum_pA}(a).
It matches well with the spectrum, which confirms that our ansatz of traveling-wave states, where the spins oscillate with positive \textit{or} negative frequency, explains the
frequency shifts in the superradiant laser.
We numerically integrate Eqs.~\eqref{eq:MF_simple} for different initial conditions and find that the spins always settle to either one of the two traveling-wave states in the long-time limit.
Noise can induce random switching between the two traveling-wave states; 
the switching rate, however, is in general exponentially suppressed in the number of spins and becomes irrelevant for a large number of spins~\cite{Fruchart_2021,Nadolny_2024}.

In active matter~\cite{Bowick_2022,Shankar_2022}, dynamical states like traveling-wave states occur via a nonreciprocal phase transition~\cite{You_2020,Fruchart_2021,Brauns_2024,Nadolny_2024}, where parity-time symmetry is broken and a time-crystalline order~\cite{Sacha_2017,Kongkhambut_2022} is induced by dynamical frustration \cite{Hanai_2024}.
Similarly, the nonreciprocal interactions between driven and undriven spins result in traveling-wave states breaking time-translation symmetry, which has direct physical relevance for the operation of superradiant lasers.
Since the cavity output closely follows the collective spin state $a = -i(2\Omega/\kappa)S^-$, the emergent dynamics of the spins imply a (positive \textit{or} negative) frequency shift of the laser, detrimental to a stable frequency reference.

The traveling-wave states are unique to the superradiant regime of lasers.
In a standard laser, where the loss rate of the cavity is small compared to the drive rate of the spins, the presence of undriven spins results in a smaller effective size of the gain medium but does not cause a frequency shift~\cite{supp}.

\mysection{Lasing transition}%
Using Eqs.~\eqref{eq:MF_simple}, a stability analysis around the incoherent state [see \cref{fig:spectrum_pA}(c)]
yields lasing above a critical fraction of driven spins~\cite{supp}
\begin{equation}
    p_{\d} > \frac{1}{2} + \frac{\gamma_+}{4V} \, .
    \label{eq:Vc_simple}
\end{equation}
Consequently, at least half of the ensemble needs to be driven for lasing to occur.
If the undriven spins were absent instead of coupled to the cavity,
the transition would be $p_\d > \gamma_+ /(2V)$, and lasing can be achieved for any $p_\d$ as long as $V$ is large enough~\cite{supp}.

\mysection{Spontaneous emission and dephasing}%
\begin{figure*}
    \centering
    \includegraphics[width = 7in]{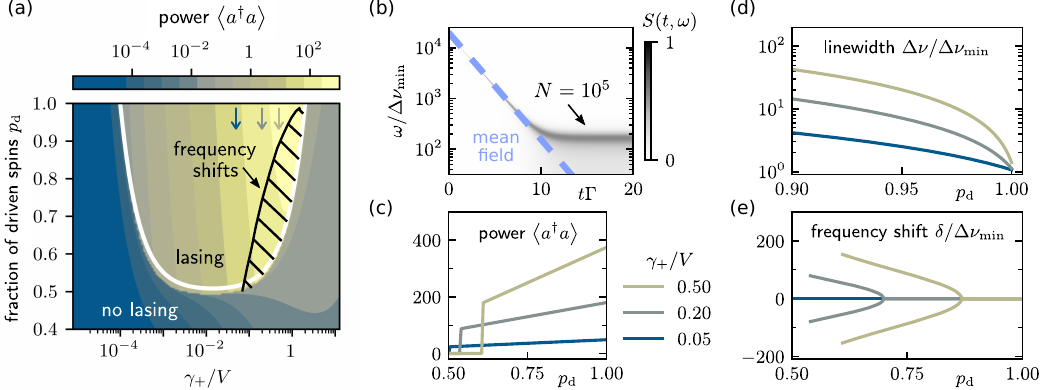}
    \caption{
    Steady-state lasing properties for $N=10^5$, $\gamma_z/V = 10^{-3}$, and $\gamma_-/V=10^{-4}$, i.e., $C = 0.2$ (other parameters as in \cref{fig:spectrum_pA}).
    (a)
    Laser power as a function of drive rate and fraction of driven spins.
    The white line shows the mean-field lasing transition. 
    The hatched region indicates the regime in which frequency shifts occur in the steady state.
    (b)
    Frequency shift as a function of time for $p_\d = 0.8$ and $\gamma_+=V$.
    The initial state at $t=0$ is the stationary state when $\gamma_-=\gamma_z=0$.
    The dashed line is the mean-field solution, which follows an exponential decay at rate $\Gamma/2=(\gamma_- + 2\gamma_z)/2$.
    The spectrum for $N = 10^5$ is shown in grayscale (arb.\ units).
    (c)
    Line cuts through (a) as marked by the three arrows in (a) on a linear scale.
    (d,e)
    Steady-state linewidth and frequency shifts in units of the minimum linewidth $\Delta \nu_\mathrm{min} = V/N$ (same legend as (c)).
    }
    \label{fig:pd}
\end{figure*}
Spontaneous emission at rate $\gamma_-$ and dephasing at rate $\gamma_z$ of each spin are included by adding 
$
\sum_{\mu = \{\d,\ud\}}
\sum_{i=1}^{N_\mu}
\left( \gamma_- \mathcal{D}[\sigma^-_{\mu,i}] + \gamma_z \mathcal{D}[\sigma^z_{\mu,i}]/2 \right)\rho
$
to the right-hand side of the master equation~\eqref{eq:master}.
As a first consequence, the lasing transition of \cref{eq:Vc_simple} changes [see~\cite{supp} for the full expression],
since the incoherent drive needs to overcome the spontaneous emission to allow for population inversion.
\Cref{fig:pd}(a) shows the steady-state laser power computed within the cumulant approximation for $N=10^5$ as a function of $\gamma_+ / V$ and $p_\d$.
The transition from small to large power is compatible with the mean-field prediction (white line).
Experimentally, $\gamma_+$ can be varied through the pump laser intensity, while the collective coupling
$V = N C \gamma_- / 2$
is set by the number of spins $N$ and the cooperativity
$C = 4\Omega^2/(\kappa\gamma_-)$.

The second consequence of spontaneous emission and dephasing is that the purity of the undriven spins is no longer conserved; their coherence decays at rate $\Gamma/2 \equiv (\gamma_- + 2\gamma_z)/2$.
As the undriven spins decohere, they become less important as antagonists to the driven spins.
The frequency shift, which arises due to the competition of alignment and antialignment, consequently decreases.
\Cref{fig:pd}(b) shows the time evolution of the frequency $\omega$.
In the mean-field limit, it exponentially decays to zero at rate $\Gamma$ (dashed line).
For finite $N$, after an initial exponential decay, a nonzero frequency shift  remains in the spectrum in the long-time limit even in the presence of spontaneous emission and dephasing.
This frequency shift vanishes for $N \rightarrow \infty$ consistent with the mean-field analysis~\cite{supp}.
The initial frequency is large compared to $\Gamma$ for typical experimental parameters~\cite{Meiser_2009} and therefore remains observable as a transient in experiments.

\mysection{Lasing properties}%
To characterize the steady-state lasing, we analyze the dependence of three key properties of the laser on the fraction of driven spins $p_\d$: the laser power, the linewidth and the frequency.
The spectral properties are obtained by fitting a double-peaked Lorentzian with linewidth $\Delta\nu$ and frequency shift $\pm\delta$ to the spectrum~\cite{supp}.
The linewidth shown in \cref{fig:pd}(d) significantly increases as the fraction of driven spins decreases, revealing a strong spectral broadening caused by the antialignment of undriven dipoles.
Taking for example $\gamma_+ = V/2$, the linewidth increases by one order of magnitude for only $3\%$ of undriven spins.
While for smaller drive rates there is less spectral broadening, the laser power is smaller, see \cref{fig:pd}(c).
Finally, as shown in \cref{fig:pd}(e), a frequency shift can occur below a critical value of $p_\d$.
This regime of nonzero frequency shifts is highlighted as the hatched region in \cref{fig:pd}(a).
It largely coincides with the regime of large powers where $\gamma_+ \approx V$.

\mysection{Conclusion}%
We have discussed a simple modification to the model of a superradiant laser that has significant consequences.
Introducing a fraction of undriven atomic spins limits the optimal operation of the laser in three ways:
an increased spectral linewidth, a shift of the lasing frequency, and a reduced power.
We explained these effects by identifying nonreciprocal interactions between driven and undriven atoms.

There are various directions to explore in further studies.
While we have focused on a continuous incoherent drive,
other proposals to achieve continuous superradiant lasing employ a beam of initially excited atoms that pass through the cavity~\cite{Kazakov_2014,Liu_2020,Tang_2022,Fama_2024}.
If a fraction of those atoms is not excited, similar dynamics as described here may occur.
Second,
one can explore distributions of incoherent drive rates other than the bimodal distribution considered here.
Third,
the effects presented in this work will similarly occur in superradiant masers operating at microwave frequencies as they are described by the same master equation~\eqref{eq:master}~\cite{Arroo_2021,Wu_2022}.
Having demonstrated practical implications of quantum nonreciprocal interactions will hopefully inspire future research of their applications in quantum technologies.

\begin{acknowledgments}
\mysection{Acknowledgments}\noindent
We acknowledge discussions with Petr Zapletal and Marco Schir\'o.
T.\,N.\ and C.\,B.\ acknowledge financial support from the Swiss National Science Foundation individual grant (No.\ 200020 200481).
M.\,B.\ acknowledges funding from the European Research Council (ERC) under the European Union’s Horizon 2020 research and innovation program (Grant agreement No.\ 101002955 - CONQUER)
\end{acknowledgments}

\pagebreak
\clearpage
\onecolumngrid
\begin{center}
\textbf{\large Supplemental material for\\``\mytitle''}
\end{center}
\newcounter{sfigure}
\renewcommand{\thefigure}{S\arabic{sfigure}}
\stepcounter{sfigure}

\newcounter{ssection}
\stepcounter{ssection}

\setcounter{page}{1}
\setcounter{equation}{0}
\makeatletter
\renewcommand{\theequation}{S\arabic{equation}}

\section{\arabic{ssection}.~Cumulant expansion}
\stepcounter{ssection}
To calculate the output power and the spectrum for large but finite $N$, we use a cumulant expansion \cite{cumulants_kubo} neglecting third and higher-order correlations.
Additionally, we exploit the permutational invariance of driven and undriven spins respectively, and the global $U(1)$-symmetry to set terms like $\expval{a}$ or $\expval{a^\dagger \sigma^+_\mu}$ equal to zero.
This leads to a closed set of eight equations,
\begin{align*}
      \frac{\mathrm{d} }{\mathrm{d} t} s^z_\mathrm{d}
      &=
      -\gamma_-(s^z_\mathrm{d}+1) -\gamma_+(s^z_\mathrm{d}-1)
      - 4 \Omega \Im[\langle  a^\dagger\sigma^-_\mathrm{d}\rangle]
      \, ,
      \\
      \frac{\mathrm{d} }{\mathrm{d} t}s^z_\mathrm{ud}
      &=
      -\gamma_-(s^z_\mathrm{ud}+1)
      - 4 \Omega \Im[\langle  a^\dagger\sigma^-_\mathrm{ud}\rangle]
      \, ,
      \\
      \frac{\mathrm{d} }{\mathrm{d} t} \expval{a^\dagger a}
      &= -\kappa \expval{a^\dagger a}
      + 2\Omega \left(
      N_\d  \Im[\expval{a^\dagger \sigma^-_\d}]+
      N_\ud \Im[\expval{a^\dagger \sigma^-_\ud}]
      \right)
      \\
      \frac{\mathrm{d} }{\mathrm{d} t} \expval{a^\dagger \sigma^-_\mathrm{d}}
      &=
      -(\gamma_++\Gamma+\kappa) \langle  a^\dagger  \sigma^-_\mathrm{d}\rangle/2 +
      i\Omega\left(
      (N_\mathrm{d}-1)\expval{\sigma^+_\mathrm{d}\sigma^-_\mathrm{d}} +
      \frac{1+s^z_\mathrm{d}}{2}+N_\mathrm{ud}\expval{\sigma^+_\mathrm{ud}\sigma^-_\mathrm{d}}
      +\expval{a^\dagger a}s^z_\mathrm{d}
      \right)
      \\
      \frac{\mathrm{d} }{\mathrm{d} t} \expval{a^\dagger \sigma^-_\mathrm{ud}}
      &=
      -(\Gamma+\kappa) \langle  a^\dagger  \sigma^-_\mathrm{ud}\rangle/2 +
      i\Omega\left(
      (N_\mathrm{ud}-1)\expval{\sigma^+_\mathrm{ud}\sigma^-_\mathrm{ud}} +
      \frac{1+s^z_\mathrm{ud}}{2}+N_\mathrm{d}\expval{\sigma^+_\mathrm{d}\sigma^-_\mathrm{ud}}
      +\expval{a^\dagger a}s^z_\mathrm{ud}
      \right)
      \, ,
      \\
      \frac{\mathrm{d} }{\mathrm{d} t}
      \expval{\sigma^+_\mathrm{d} \sigma^-_\mathrm{d}}
      &=
      -(\gamma_++\Gamma)\expval{\sigma^+_\mathrm{d} \sigma^-_\mathrm{d}}
	  +2\Omega s^z_\mathrm{d} \Im[\langle  a^\dagger \sigma^-_\mathrm{d}\rangle]
      \, ,
      \\
      \frac{\mathrm{d} }{\mathrm{d} t}
      \expval{\sigma^+_\mathrm{ud} \sigma^-_\mathrm{ud}}
      &=
      -\Gamma\expval{\sigma^+_\mathrm{ud} \sigma^-_\mathrm{ud}}
      +2\Omega s^z_\mathrm{ud} \Im[\langle  a^\dagger \sigma^-_\mathrm{ud}\rangle]
      \\
      \frac{\mathrm{d} }{\mathrm{d} t}
      \expval{\sigma^+_\mathrm{d} \sigma^-_\mathrm{ud}}
      &=
      -(\gamma_+/2+\Gamma)\expval{\sigma^+_\mathrm{d} \sigma^-_\mathrm{ud}}
      +i\Omega \left(
        s^z_\mathrm{ud} \expval{a^\dagger \sigma^-_\mathrm{d}}^*
       -s^z_\mathrm{d} \expval{a^\dagger \sigma^-_\mathrm{ud}}
      \right)
      \, .
\end{align*}
We use the Julia package QuantumCumulants.jl~\cite{Plankensteiner2022quantumcumulantsjl} to integrate these equations numerically.

\section{\arabic{ssection}.~Spectrum}
\stepcounter{ssection}
The spectrum is computed within the cumulant expansion approximation using the quantum regression theorem~\cite{Breuer}.
By factorizing $\expval{\sigma^z_\mu(t+\tau) a^\dag(t+\tau) a(t)} \approx s^z_\mu(t+\tau) \expval{a^\dag(t+\tau) a(t)}$, the two-time correlations evolve according to 
\begin{equation}
    \frac{\mathrm{d}}{\mathrm{d}\tau}
    \mathbf{c}(t,\tau)
    =
    M(t+\tau)
    \mathbf{c}(t,\tau)
    \, , \quad 
    \mathbf{c}(t,\tau) =
    \begin{pmatrix}
        \expval{a^\dag(t+\tau) a(t)}
        \\
        \expval{\sigma^+_\d(t+\tau) a(t)}
        \\
        \expval{\sigma^+_\ud(t+\tau) a(t)}
    \end{pmatrix}
    \, , \quad 
    M(t+\tau) = \begin{pmatrix}
        -\kappa / 2 & iN_\d \Omega & iN_\ud \Omega
        \\
        -i\Omega s^z_\mathrm{d}(t+\tau) & -(\Gamma + \gamma_+)/2 & 0
        \\
        -i\Omega s^z_\mathrm{ud}(t+\tau) & 0 & -\Gamma /2 
    \end{pmatrix}
    \, .
    \label{eq:supp_two_time}
\end{equation}
In the steady state $t\rightarrow \infty$, where $s^z_\mu$ obtains a constant value, the matrix $M$ is time independent, and the steady-state spectrum can be calculated using the Laplace transform,
\begin{equation}
    S(\omega) = 
    \lim_{t\rightarrow\infty}
    \int_{-\infty}^\infty \mathrm{d} \tau \expval{a^\dag(t+\tau)a(t)} e^{i\omega\tau}
    = 
    2\Re{
    \left[
    (i\omega - M)^{-1} \mathbf{c}_\mathrm{ss}
    \right]_1}
    \, .
\end{equation}
Here, the subscript $1$ refers to the first component of the vector, and $\mathbf{c}_\mathrm{ss}=\lim_{t\rightarrow\infty} \mathbf{c}(t,0)$.

The spectrum in \cref{fig:pd}(b) can be calculated by a discrete Fourier transform of solutions of \cref{eq:supp_two_time} in consecutive time intervals.
By using that $s^z_\mu$ changes at rate $\Gamma$ during the transient, which is small compared to the typical frequency $\omega$, we set $s^z_\mu(t+\tau) = s^z_\mu(t)$ constant for each time interval and use the Laplace transform at each time $t$ to approximately calculate the spectrum,
\begin{equation}
    S(t,\omega) = 
    \int_{-\infty}^\infty \mathrm{d} \tau \expval{a^\dag(t+\tau)a(t)} e^{i\omega\tau}
    \approx
    2\Re{
    \left[
    (i\omega - M(t)^{-1} \mathbf{c}(t,0)
    \right]_1}
    \, .
\end{equation}

\section{\arabic{ssection}.~Influence of cavity-spin detuning}
\stepcounter{ssection}
We discuss the effect of a nonzero detuning between cavity frequency $\omega_c$ and spin frequency $\omega_s$.
The master equation in the laboratory frame is
\begin{equation}
    \dot \rho
    =
    -i[H,\rho] + \kappa \mathcal{D}[a]\rho
    +\gamma_+ \sum_{i=1}^{N_\d} \mathcal{D}[\sigma^+_{\d,i}]\rho 
    \, , 
	\label{eq:master_supp}
% \end{equation}
% with Hamiltonian~\cite{Meiser_2009}
% \begin{equation}
\quad
    H = \frac{\omega_s}{2} S^z + \omega_c a^\dag a + \Omega(a^\dag S^- + a S^+)
    \, .
\end{equation}
The collective spin operators are as before $S^{\pm,z} = \sum_{i=1}^{N_{\d}} \sigma_{\d,i}^{\pm,z} + \sum_{i=1}^{N_{\ud}} \sigma_{\ud,i}^{\pm,z}$.
In the frame where both spins and cavity rotate with the spin frequency $\omega_s$, the Hamiltonian becomes
\begin{equation}
    H = (\omega_c-\omega_s) a^\dag a + \Omega(a^\dag S^- + a S^+)
    \, .
\end{equation}
The cavity-spin detuning $\omega_c-\omega_s$ results in a frequency shift of the laser.
% presenting a challenge for the stable operation of a superradiant laser.
In the bad-cavity limit, the frequency shift is smaller than the detuning by a factor of approximately $\Gamma/\kappa \sim 10^{-3}-10^{-5}$ and therefore does not limit the stability of a superradiant laser~\cite{Bohnet_2012,Norcia_2016}. 
% Therefore, stabilizing the cavity to a $1 - 100\,\mathrm{Hz}$ is sufficient and achievable.
Note the qualitative difference between the frequency shift due to nonzero detuning and the frequency shifts due to nonreciprocal interactions. 
A nonzero detuning induces a unique and deterministic frequency shift. 
The nonreciprocal interactions, however, result in one of two possible frequency shifts through spontaneous symmetry breaking.

\begin{figure}
    \centering
    \includegraphics[width=5.6in]{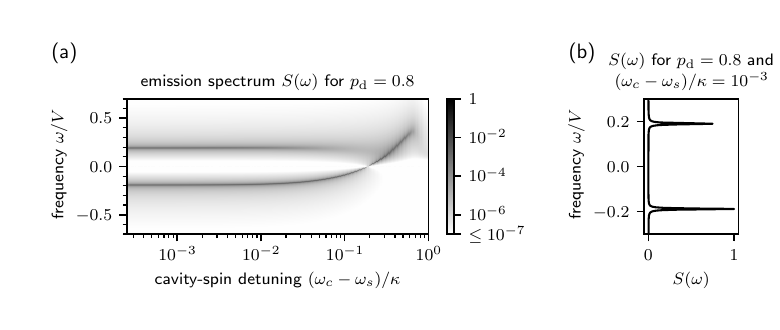}
    \caption{
    Influence of detuning $\omega_c-\omega_s$ between cavity and spins on the frequency shift.
    Parameters are the same as in \cref{fig:spectrum_pA} with $p_\d = 0.8$.
    (a)~Cavity emission spectrum $S(\omega)$ (arb.\ units) as a function of cavity-spin detuning $\omega_c-\omega_s$.
    The frequency $\omega$ is measured relative to the bare spin frequency.
    For negative detuning, $\omega_c-\omega_s < 0$, the same result is obtained but mirrored at $\omega = 0$.
    (b)~Line cut through (a) at $\omega_c-\omega_s = 10^{-3} \kappa$ 
    %for 
    on a linear scale highlighting the asymmetry in the two peaks.
    }
    \label{fig:supp2}
\end{figure}

The influence of the detuning $\omega_c-\omega_s$ between cavity and spins on the traveling-wave states, and correspondingly the frequency shifts, is shown in \cref{fig:supp2}.
Since the parity-time symmetry of the master equation is only exact when $\omega_c-\omega_s = 0$ (see also Ref.~\cite{Nadolny_2024}), the detuning explicitly breaks the symmetry between the two frequencies of the traveling-wave states.
Therefore, one of the two traveling-wave states is more likely to occur, which is indicated by the asymmetry in the peaks, see \cref{fig:supp2}(b).
However, for experimentally relevant values of the detuning, $\omega_c-\omega_s  \ll \kappa$, the influence of the detuning on the frequency shifts induced by nonreciprocal interactions is negligible.
When the detuning is comparable to the cavity dissipation rate, the coherent spin-spin interactions mediated by the cavity result in rich physics beyond the scope of the present work~\cite{Norcia_2018b}.

\section{\arabic{ssection}.~Comparison to standard laser}
\stepcounter{ssection}

The mean-field equations of the master equation Eq.~\eqref{eq:master} in the presence of the cavity field $\alpha = \expval{a}$ are
\begin{equation}
    \begin{split}
      \frac{\mathrm{d} }{\mathrm{d} t} s^z_\mathrm{d}
      &=
      -\gamma_-(s^z_\mathrm{d}+1) -\gamma_+(s^z_\mathrm{d}-1)
      + 4 \Omega \Im[\alpha s^+_\mathrm{d}]
      \, ,
      \\
      \frac{\mathrm{d} }{\mathrm{d} t}s^z_\mathrm{ud}
      &=
      -\gamma_-(s^z_\mathrm{ud}+1) 
      + 4 \Omega \Im[\alpha s^+_\mathrm{ud}]
      \, ,
      \\
      \frac{\mathrm{d} }{\mathrm{d} t} s^+_\mathrm{d}
      &=
      -(\gamma_++\gamma_-+2\gamma_z)    s^+_\mathrm{d}/2 -
      i\Omega \alpha^* s^z_\mathrm{d}
      \, ,
      \\
      \frac{\mathrm{d} }{\mathrm{d} t}  s^+_\mathrm{ud}
      &=
      -(\gamma_-+2\gamma_z)    s^+_\mathrm{ud}/2 -
      i\Omega \alpha^* s^z_\mathrm{ud}
      \, ,
      \\
      \frac{\mathrm{d} }{\mathrm{d} t} \alpha
      &= -\kappa \alpha/2 - iN\Omega s^-
      \, .
      \end{split}
      \label{eq:MF}
\end{equation}
They result in Eqs.~\eqref{eq:MF_simple} after the adiabatic elimination of the cavity field, $\alpha \rightarrow -2iN\Omega s^-/\kappa$.

In a standard laser, the cavity is a `good' cavity with a decay rate $\kappa$ that is small compared to the incoherent gain and damping of the atoms.
In this case, one can adiabatically eliminate the spin degrees of freedom in \cref{eq:MF} by setting $\dot s^z_{\mu} = \dot s^+_{\mu} = 0$ to obtain
\begin{equation}
    \frac{\mathrm{d} }{\mathrm{d} t}  \alpha
    =
    \alpha
    \left(
    -\frac{\kappa}{2}
    + 2N p_\mathrm{d}
    \frac{\gamma_+ - \gamma_-}{\tilde \Gamma_\mathrm{d}^2/\Omega^2+8\abs{\alpha}^2}
    - 2N p_\mathrm{ud} 
    \frac{\gamma_- }{\tilde \Gamma_\mathrm{ud}^2/\Omega^2+8\abs{\alpha}^2}
    \right)
    \, ,
\end{equation}
where we have defined
$\tilde \Gamma_\mathrm{d}^2 =
(\gamma_- + \gamma_+)(\gamma_- + \gamma_+ + 2\gamma_z)$ and
$\tilde \Gamma_\mathrm{ud}^2 = \gamma_- ( \gamma_- + 2\gamma_z)$.
Since the term in brackets on the right-hand side is real-valued, there are no oscillations and no traveling-wave states.
The lasing transition (obtained by expanding around $\alpha = 0$) is
\begin{equation}
p_\d  \frac{\gamma_+ - \gamma_-}{\tilde\Gamma_\mathrm{d}^2 } - 
p_\ud \frac{ \gamma_-}{\tilde\Gamma_\mathrm{ud}^2}
>
\frac{ \kappa}{4N\Omega^2} = \frac{1}{2V}
\, .
\end{equation}
For any nonzero $p_\d$, the drive rate $\gamma_+$ can be increased sufficiently to obtain lasing.
This contrasts the behavior of the superradiant laser, where no lasing can be obtained for $p_\d <1/2$, see \cref{eq:Vc_simple} or \cref{eq:Vc_full}.

\section{\arabic{ssection}.~Mean-field stability analysis}
\stepcounter{ssection}
The mean-field equations including decay at rate $\gamma_-$ and dephasing at rate $\Gamma = \gamma_- + 2\gamma_z$ are
\begin{equation}
    \begin{split}
        \frac{\mathrm{d}}{\mathrm{d}t} {s}_{\d}^+ 
        &= V s^+ s^z_{\d} 
        - \left(\Gamma + \gamma_+ \right) s^+_{\d}/2 
        \, ,
        \\
        \frac{\mathrm{d}}{\mathrm{d}t} {s}_{\ud}^+ 
        &= V s^+ s^z_{\ud} 
        - \Gamma s^+_{\ud}/2 
        \, ,
        \\
        \frac{\mathrm{d}}{\mathrm{d}t} {s}_{\d}^z 
        &= -4V \Re[{s}_{\d}^- s^+]
        - \gamma_- (1 + {s}_{\d}^z) + \gamma_+ (1 - {s}_{\d}^z) 
        \, ,
        \\
        \frac{\mathrm{d}}{\mathrm{d}t} {s}_{\ud}^z 
        &= -4V \Re[{s}_{\ud}^- s^+]
        - \gamma_- (1 + {s}_{\ud}^z) 
        \, ,
    \end{split}
    \label{eq:MF_full}
\end{equation}
where again $s^+ = p_\d s^+_\d+p_\ud s^+_\ud$.
The fixed point, which characterizes the incoherent state, is $s^+_{\d,\ud} = 0$, $s^z_\d = (\gamma_+-\gamma)/(\gamma_++\gamma)$, and $s^z_\ud = -1$.
From a stability analysis around this fixed point, we obtain the lasing transition as
\begin{equation}
    p_\d >
    \frac{1}{2}
    \left(1+\frac{\gamma_-}{\gamma_+} \right)
    \left(1+\frac{\Gamma}{V}+\frac{\gamma_+}{2V} \right)
    \, ,
    \label{eq:Vc_full}
\end{equation}
which reduces to \cref{eq:Vc_simple} when $\gamma_- = \gamma_z = 0$, and defines the white line shown in \cref{fig:pd}(a).

If the undriven spins do not couple to the cavity (setting $V=0$ in the second and fourth line of Eqs.~\eqref{eq:MF_full}), the critical coupling is
\begin{equation}
    p_\d > \frac{(\gamma_- + \gamma_+) (\Gamma + \gamma_+)}{2 (\gamma_+-\gamma_-) V}
    \xrightarrow{\gamma_- = \Gamma = 0} \frac{\gamma_+}{2V}
    \, .
\end{equation}
The lasing threshold is smaller in this case compared to the threshold set by \cref{eq:Vc_full}.
Therefore, the undriven spins increase the lasing threshold more when they are coupled to the cavity compared to when they are not coupled.

\section{\arabic{ssection}.~Lorentzian fits}
\stepcounter{ssection}
To obtain the linewidth and the frequency shift in Figs.~\ref{fig:pd}(d,e), we fit the sum of two Lorentzian distributions,
\begin{equation}
    \frac{A}{\pi} \left( \frac{\Delta \nu}{\Delta \nu^2 + (\omega - \delta)^2} + \frac{\Delta \nu}{\Delta \nu^2 + (\omega + \delta)^2} \right)\, ,
\end{equation}
with amplitude $A$ and linewidth $\Delta \nu$, displaced by $\pm \delta$, to the spectrum.
The fitted linewidth and frequency shift agree well with the real and imaginary parts of the eigenvalues of matrix $M$ of \cref{eq:supp_two_time}.

\section{\arabic{ssection}.~Dependence of frequency shift on system size}
\begin{figure}[H]
    \centering
    \includegraphics[width=\linewidth]{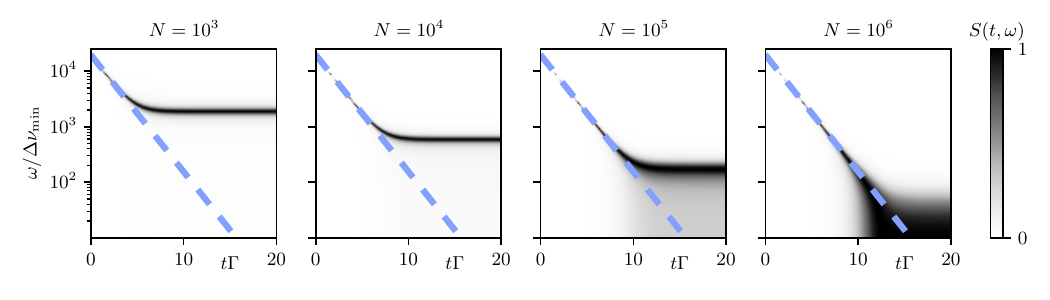}
    \caption{Time dependence of the frequency shift in the spectrum for different values of $N$. 
    The spectrum is normalized to a maximum value of $1$ for each time $t$.
    The dashed line is the mean-field prediction of the frequency.
    The initial state at $t=0$ is the stationary state when $\gamma_-=\gamma_z=0$.
    Parameters as in \cref{fig:pd}(b) with $N$ specified for each panel.
    }
    \label{fig:supp_Nscaling}
\end{figure}

\stepcounter{ssection}
\Cref{fig:supp_Nscaling} shows the time evolution of the frequency shift for different values of $N$.
For $N\rightarrow \infty$, the spectrum approaches the mean-field prediction (dashed line).
The cumulant expansion thus predicts that the frequency shift vanishes for $t\to\infty$ in the thermodynamic limit consistent with the mean-field analysis.
\stepcounter{sfigure}

\end{document}